\documentclass{cjaa}                   

\usepackage{graphicx}                   
\input{epsf.sty}                        
\input{psfig.sty}                       

\newcommand{\beq}{\begin{equation}}
\newcommand{\eeq}{\end{equation}}
\newcommand{\beqn}{\begin{eqnarray}}
\newcommand{\eeqn}{\end{eqnarray}}
\newcommand{\pd}{\partial}

\begin{document}

   \title{Particle acceleration in astrophysical shear flows}

   \volnopage{Vol.0 (200x) No.0, 000--000}      
   \setcounter{page}{1}          

   \author{Frank M. Rieger
      \mailto{}
    \and Peter Duffy
      }
   \offprints{Frank Rieger}                   

   \institute{Department of Mathematical Physics, University College Dublin, Belfield, Dublin 4,
              Ireland\\
             \email{frank.rieger@ucd.ie}
          }

   \date{Received~~2004; accepted~~2004}

   \abstract{We consider the acceleration of particles due to a velocity shear in relativistic 
    astrophysical flows. The basic physical picture and the formation of power law momentum spectra
    is discussed for a non-relativistic velocity field using a microscopic approach. We identify 
    possible sites for shear acceleration in relativistic astrophysical jets and analyze their 
    associated acceleration timescales. It is shown in particular that for a mean scattering time 
    $\tau$ scaling with the gyro-radius, the acceleration timescale for gradual shear scales in the 
    same manner as the synchrotron cooling timescale so that losses may no longer be able to stop 
    the acceleration once it has started to work efficiently. Finally, the possible role of shear 
    acceleration is discussed with reference to the relativistic jet in the quasar 3C~273.
    }

   \authorrunning{F.M. Rieger \& P. Duffy}            
   \titlerunning{Particel acceleration in astrophysical shear flows}

   \maketitle

\section{Introduction}\label{sect:intro}
Collimated relativistic outflows are observed in a variety of astrophysical environments, ranging 
from Active Galactic Nuclei (AGN), $\mu$-Quasars and neutron stars in binary systems to Gamma-Ray 
Bursts (e.g., Zensus~1997; Mirabel \& Rodriguez~1999; Fender et al.~2004). There is growing evidence 
today that in all these sources the relativistic outflows, or at least those parts observed in the 
radio band, are essentially launched from an accretion disk (e.g., Livio~1999; Marscher et al.~2002; 
Fender et al.~2004). In several sources, monitoring of their small-scale radio structures has 
revealed outflowing components at (apparently) high superluminal velocities suggesting that the bulk 
plasma in these jets moves at relativistic speeds and small viewing angles towards the observer. 
Theoretical models based on the assumption of a simple, one-dimensional velocity structure have thus 
allowed valuable insights into the emission properties of such jets. In general however, such models 
are only adequate as a first approximation, as real jets are likely to be characterized by a significant 
shear in their flow velocity field. Using theoretical and observational arguments relevant for AGN jets 
for example, we show below that at least three different shear flow scenarios may be distinguished. 
In the presence of such shear flows, particle acceleration is likely to occur: In their pioneer work 
on the particle transport in non-relativistic gradual shear flows for example, Berezhko \& Krymskii 
(1981) (cf. also Berezhko~1981) showed that (steady-state) power law particle momentum spectra $f(p) 
\propto p^{-(3+\alpha)}$ may be formed if the scattering time follows a power law $\tau \propto 
p^{\alpha}$, $\alpha > 0$. 
Later on, Earl, Jokipii \& Morfill~(1988) re-derived Parker's transport equation for non-relativistic 
gradual shear flows, but now augmented by new terms describing viscous (shear) and inertial effects. 
A relativistic generalization was achieved by Webb~(1989) (cf. also Webb~1985 and Webb et al.~1994) 
based on a mixed-frame approach, where the scattering operator is evaluated in the comoving frame, and
applied to the cosmic ray transport in our galaxy. Particle acceleration in non-gradual relativistic 
shear flow has been considered by Ostrowski (1990,~1998,~2000; cf also Stawarz \& Ostrowski~2002) based 
on Monte Carlo simulations, showing that very flat particle momentum spectra may be possible. More 
recently, Rieger \& Mannheim~(2002) have studied the acceleration of particles in rotating and shearing 
flows with application to relativistic AGN jets. Based on the insights gained from these works, shear 
acceleration in astrophysical jets appears in some respects unavoidable. Whether shear-accelerated 
particles may significantly contribute to the emission in a given energy band seems thus not a matter 
of occurrence but simply a matter of efficiency of the underlying shear acceleration mechanism.

\section{A simplified approach to shear acceleration}
Shear acceleration is based on the idea that energetic particles may gain energy by scattering 
off systematically moving small-scale magnetic field irregularities. These scattering centres 
(SCs) are thought to be embedded in a collisionless (gradual) shear flow such that the velocity 
of a certain SC corresponds to the local flow velocity, i.e. in contrast to second order Fermi 
acceleration a random motion of SCs is neglected. Hence the acceleration process essentially draws 
on the kinetic energy of the background flow. Scattering of particles is assumed to occur in such 
a way that the particle momentum is randomized in direction and its magnitude is conserved in the 
local comoving fluid/scattering frame. Obviously, if there is no velocity shear (or rotation) 
present and the flow is not diverging, particles will neither gain energy nor momentum merely due 
to scattering. However, in the presence of a velocity shear in the flow, the momentum of a particle 
travelling across the shear changes, so that a net increase may occur. The physical argument perhaps 
becomes best transparent in the microscopic picture for a non-relativistic continuous shear flow 
(cf., Jokipii \& Morfill~1990) with velocity profile given by $\vec{u} = u_z(x)\,\vec{e}_z$. Consider 
a particle with velocity vector $\vec{v}=(v_x,v_y,v_z)$, relativistic mass $m$ and initial momentum 
$\vec{p}_1$ relative to local flow frame. Within a scattering time $\tau$ (assumed to be independent
of momentum) the particle will travel a distance $\delta x = v_x\,\tau$ across the shear, so that 
for a gradual shear flow the flow velocity will have changed by an amount $\delta \vec{u} = \delta u\,
\vec{e_z}$, where $\delta u = (\partial u_z/\partial x)\,\delta x$. Contenting ourselves with a 
Galilean transformation for the non-relativistic flow speeds involved, the particle's momentum 
relative to the flow will have changed to $\vec{p}_2 = \vec{p}_1 + m\,\delta \vec{u}$, i.e. 
\beq\label{trafo}
    p_2^2 = p_1^2 + 2\,m\,\delta u\,\,p_{1,z} + m^2\,(\delta u)^2\;.
\eeq As the next scattering event preserves the magnitude of the particle momentum relative to 
the local flow speed, the particle magnitude will have this value in the local flow frame and hence 
a net increase in momentum may occur with time. By using spherical coordinates, defining $\mbox{$
<\Delta p/\Delta t>$} =  \mbox{$2\,<(p_2 - p_1)>/\tau$}$ as the average rate of momentum 
change and $\mbox{$<\Delta p^2/\Delta t>$} = \mbox{$2\,<(p_2 - p_1)^2>/\tau$}$ as the average rate 
of momentum dispersion, it can be shown that averaging over solid angles assuming a nearly 
isotropic particle distribution results in (cf., Jokipii \& Morfill~1990)
\beqn
   \left<\frac{\Delta p^2}{\Delta t}\right> & = & \frac{2}{15}\,p^2
   \left(\frac{\pd u_z}{\pd x}\right)^2 \tau \;\\
   \left<\frac{\Delta p}{\Delta t}\right> & = & \frac{4}{15}\,p
   \left(\frac{\pd u_z}{\pd x}\right)^2 \tau \;,\label{p_dot}
\eeqn both being related by 
\beq 
    \left<\frac{\Delta p}{\Delta t}\right> = \frac{1}{2\,p^2}\,\frac{\pd}{\pd p}
         \left(p^2 \left<\frac{\Delta p^2}{\Delta t}\right>\right)\,.
\eeq Note that a momentum-dependent scattering time obeying a power law of the form $\tau \propto 
p^{\alpha}$ may be accommodated by replacing $4 \rightarrow (4+\alpha)$ in Eq.~(\ref{p_dot}). We 
may then gain further insights into possible particle momentum spectra under viscous shear energy 
changes by studying the (simplified) steady state transport equation for the phase space distribution 
function $f(p)$
\beq\label{transport}
  \frac{1}{p^2}\,\frac{\pd}{\pd p} \left(p^2 \left<\frac{\Delta p}{\Delta t}\right> f(p)\right) 
  - \frac{1}{2\,p^2}\frac{\pd^2}{\pd p^2} \left(p^2 \left<\frac{\Delta p^2}{\Delta t}\right> f(p) 
  \right) = Q\, \delta(p-p_0)\,,
\eeq assuming particles to be injected with momentum $p_0$ and $\alpha > 0$. 
Solving Eq.~({\ref{transport}) one immediately obtains
\beq
     f(p) \propto p^{-\,(3 + \alpha)}\,\, H(p-p_0)\,,
\eeq where $H(p)$ is the Heaviside step function. Note that for a mean scattering time scaling 
with the gyro-radius (Bohm case), i.e. $\tau \propto p$, $\alpha =1$, this gives $f(p) \propto 
p^{-4}$ and thus a power law particle number density $n(p) \propto p^{-2}$ translating into a 
synchrotron emissivity $j_{\nu} \propto \nu^{-1/2}$. For a Kolmogorov-type ($\alpha =1/3$) or 
Kraichnan-type ($\alpha=1/2$) scaling on the other hand, much flatter spectra are obtained.

\section{Possible velocity shear sites in astrophysical jets}
In general, at least three different shear scenarios may be distinguished (cf. Rieger \& Duffy~2004):
\subsection{Gradual velocity shear parallel to the jet axis}
Observationally, we have strong direct evidence for internal jet stratification (e.g. such as a fast 
moving inner spine and a slower moving boundary layer) on the pc-scale in sources such as 3C353, M87, 
Mkn~501 or PKS 1055+018 (e.g., Swain et al.~1998; Attridge et al.~1999; Perlman et al.~1999; Edwards 
et al.~2000). Moreover, there is a multitude of indirect, phenomenological evidence for internal jet 
stratification including the requirements for the unification of BL Lacs and FR~I (e.g., Chiaberge et 
al.~2000) or results from hydrodynamical jet simulations (e.g., Aloy et al.~2000; Gomez~2002). In 
particular, Laing et al.~(1999) have argued recently that the intensity and polarization systematics 
in kpc-scale FR~I jets are suggestive of a radially (continuously) decreasing velocity profile $v_z(r)$. 
\subsection{Non-gradual velocity shear parallel to the jet axis}
In general, a gradual shear analysis may be justified if the flow velocity changes continuously and 
the particle mean free path $\lambda$ is much smaller than the width of the shear region. However, 
when a particle becomes so energetic that its mean free path becomes larger than the width of the 
velocity transition layer, it will essentially experience a non-gradual shear flow discontinuity. 
Ostrowski~(1990,1998) for example, has convincingly argued that the jet side boundary between the 
relativistic jet interior and its ambient medium may represent a natural realization of such a 
non-gradual (discontinuous) relativistic velocity shear.
\subsection{Gradual velocity shear across the jet}
Astrophysical jets are also likely to exhibit a significant velocity shear perpendicular to their 
jet axis ('transversal shear'). In particular, several independent arguments suggest that 
astrophysical jets may be characterized by an additional rotational velocity component: The strong 
correlation between the disk luminosity and the bulk kinetic power in the jet (e.g., Rawlings \& 
Saunders~1991), the successful applications of jet-disk symbiosis models (e.g., Falcke \& 
Biermann~1995) and the observational evidence for a disk-origin of jets (e.g., Marscher et al.~2002)
or periodic variability (e.g., Camenzind \& Krockenberger~1992) suggest that a significant amount 
of accretion energy and hence rotational energy of the disk is channeled into the jet (due to angular 
momentum conservation). Moreover, internal jet rotation is generally implied in theoretical MHD 
models of jets as magnetized disk winds (e.g., Vlahakis \& Tsinganos~1998; Sauty et al.~2002). 
Finally, direct observational support for internal jet rotation has been recently established for 
stellar jets (e.g., Coffey et al.~2004). Based on such arguments, Rieger \& Mannheim~(2002) have 
analyzed the acceleration of particles in rotating and shearing, relativistic AGN jets. Their analysis 
confirmed the results of Berezhko \& Krymskii~(1981) for the case of non-relativistic Keplerian rotation, 
but also showed that inclusion of centrifugal effects may lead to a flattening of the spectra.

\section{On the efficiency of shear acceleration}
Rieger \& Duffy~(2004) have recently studied the acceleration potential for the different
shear scenarios distinguished above by estimating their associated acceleration timescales.
For a gradual shear flow parallel to the jet axis (case 3.1.) with a velocity profile 
decreasing linearly from relativistic to non-relativistic speeds over a scale $\Delta r$,
they found a minimum acceleration timescale
\beq
   t_{\rm grad,||} \sim \frac{3}{\lambda}\,\frac{(\Delta r)^2}{c\,\gamma_b(r)^4}\,,
\eeq where $\gamma_b(r) >1 $ is the (position-dependent) bulk Lorentz factor of the flow 
and $\lambda$ the particle mean free path. By comparing the acceleration with the synchrotron 
cooling timescale for parameters appropriate for pc-scale radio jets, it can be shown that 
efficient electron acceleration is quite restricted and only possible if $\Delta r$ is very 
small, while proton acceleration is much more favourable since $\lambda_{\rm proton} \gg 
\lambda_{\rm electron}$. Perhaps most interestingly however, for a particle mean free path 
scaling with the gyro-radius, i.e. $\lambda \propto \gamma$, both the acceleration and the 
cooling timescale have the same dependency on the particle Lorentz factor $\gamma$. In such 
a case losses are no longer able to stop the acceleration process once it has started to work 
efficiently. Particle escape, when the $\lambda$ becomes larger than the width of the jet, or 
cross-field diffusion however, may then still represent important constraints on the maximum 
attainable energy.\\
For a non-gradual, relativistic velocity jump (case 3.2.) on the other hand, the minimum 
acceleration timescale is of order (cf. Ostrowski~1990,~1998)
\beq
   t_{\rm acc} \sim 10\, \frac{r_g}{c} \quad\quad {\rm provided}\quad r_g > \Delta r\,,
\eeq where $r_g$ denotes the gyro-radius of the particle and $\Delta r$ the width of the
transition layer, so that the acceleration timescale in this case is proportional to $\gamma$. 
Due to the condition $r_g > \Delta r$, efficient acceleration of electrons is unlikely given 
their associated rapid radiation losses. However, the acceleration of protons may be well 
possible until their particle mean free path becomes larger than the width of the jet.\\
Finally, for a shear flow with relativistic $v_z$ and azimuthal Keplerian rotation profile 
$\Omega(r) \propto (r_{\rm in}/r)^{3/2}$ for example (case 3.3.), the acceleration timescale 
obeys 
\beq
  t_{{\rm grad},\perp} \propto \frac{1}{\lambda}\,\left(\frac{r}{r_{\rm in}}\right)^3\,,
\eeq assuming the flow to be radially confined to $r_{\rm in} \leq r \leq r_j$ where $r_j$ is
the jet radius. In general, efficient particle acceleration requires a region with significant 
rotation. For such a case the higher energy emission will naturally be concentrated closer to 
the axis (i.e. towards smaller radii). Again however, electron acceleration is usually very 
restricted (i.e. only possible for $r \sim r_{\rm in}$), while proton acceleration is much more 
favourable (i.e. possible up to $r \sim r_j$). If $\lambda$ scales with the gyro-radius, the 
acceleration and the cooling timescale will again have the same dependency on $\gamma$ and thus 
losses will not longer be able to stop the acceleration process once it has started to work 
efficiently.

\section{Possible applications and relevance}
Observational and theoretical evidence suggest that astrophysical jets are characterized by 
a significant velocity shear. While internal jet rotation for example, is likely to be present 
at least in the initial parts of the jet, a significant longitudinal velocity shear (parallel to 
the jet axis) prevailing all along the jet might be expected for most sources. In general, the shear 
acceleration mechanism analyzed above may thus operate over many length scales. Thus, in contrast 
to shock acceleration, shear acceleration is usually not constrained to localized regions in the 
jets. Besides the possibility that shear acceleration may account for a second population of
energetic particles in the jet (e.g. in addition to shock-accelerated ones), shear acceleration 
might be of particular relevance for our understanding of the continuous radio and optical 
emission observed from several sources (e.g., Meisenheimer et al.~1997; Scarpa et al.~1999;
Jester et al.~2001). In the case of the quasar 3C~273 ($z=0.158$) for example, the optical 
spectral index is found to vary only very smoothly along the (large-scale) jet with no signs 
of strong synchrotron cooling at any location in the jet, e.g. between knots, contrary to
expectations from shock acceleration scenarios (Jester et al.~2001), suggesting the need 
of a continuous re-acceleration mechanism (cf. also Meisenheimer et al.~1997). The jet in 3C~273 
appears to be highly relativistic even on kpc-scales with a typical Doppler factor $D \sim 5$ 
(cf. Sambruna et al~2001) so that longitudinal shear acceleration of particles may probably work 
efficiently nearly all along the jet. Moreover, there is also evidence for helical bulk motion 
in the large-scale jet (e.g., Bahcall et al.~1995) and internal jet helicity (e.g., oscillating
ridge line, double helical pattern, periodicity) on the VLBI mas-scale and below (e.g., Krichbaum 
et al.~2000, Camenzind \& Krockenberger~1992), suggesting that particle acceleration due to 
internal jet rotation (cf. Rieger \& Mannheim~2002) may contribute on pc-scale and perhaps even 
on larger scales.


\begin{acknowledgements}
FMR acknowledges support by a Marie-Curie Individual Fellowship (MCIF-2002-00842). 
Discussions with V.~Bosch-Ramon and K.~Mannheim are gratefully acknowledged.
\end{acknowledgements}

\label{lastpage}

\end{document}